**Manuscript title:** A preliminary study of liver fat quantification using reported longitudinal ultrasound speed of sound and attenuation parameters

**Authors' names:** Juvenal Ormachea[1], Kevin J. Parker[1]

[1]Department of Electrical and Computer Engineering, University of Rochester, Rochester, NY, USA

**Corresponding author:** Kevin J. Parker
University of Rochester
Computer Studies Building 724
Box 270231
Rochester, NY 14627-0231
Phone: (585)275-3294
Fax: (585)275-2073
Email: kevin.parker@rochester.edu

**Abstract**

The quantification of liver fat as a diagnostic assessment of steatosis remains an important priority for noninvasive imaging systems. We derive a framework in which the unknown fat volume percentage can be estimated from a pair of ultrasound measurements. The precise estimation of ultrasound speed of sound and attenuation within the liver are shown to be sufficient for estimating fat volume assuming a classical model of the properties of a composite elastic material. In this model, steatosis is represented as a random dispersion of spherical fat vacuoles with acoustic properties similar to those of edible oils. Using values of speed of sound and attenuation from the literature where normal and steatotic livers were studied near 3.5 MHz, we demonstrate agreement of the new estimation method with independent measures of fat. This framework holds the potential for translation to clinical scanners where the two ultrasound measurements can be made and utilized for improved quantitative assessment of steatosis.

# INTRODUCTION

Two historic circumstances have motivated a renewed effort to quantify the amount of fat in liver, and thereby assess the progression of steatosis with some degree of accuracy. The first motivating factor is the increasing prevalence of fatty liver across the globe, including in younger individuals (Browning et al. 2004; Diehl and Day 2017). The second factor is the increasing capabilities of ultrasound scanners to measure parameters related to compression and shear wave phenomena. These have resulted in a number of techniques and metrics which are reviewed in recent publications (Pirmoazen et al. 2020; Ferraioli et al. 2021). Generally speaking, a traditional approach is to correlate a single parameter, such as the speed of sound for example, against a steatosis grade in some population as defined by some independent standard. Histology assessments from liver biopsy has been typically used as a gold standard but these are frequently scored with subjective ratings. More recently, magnetic resonance imaging techniques have gained acceptance as a more quantitative and reliable standard (Caussy et al. 2018). However, the single-parameter correlations suffer from imprecision in measurements, biological variability, and the inevitable presence of confounding cofactors that are unmeasured yet can influence the parameter being studied. Multiple parameters measured simultaneously can improve the assessment of the degree of steatosis (Baek et al. 2020a; Baek et al. 2020b; Baek et al. 2020c; Basavarajappa et al. 2020; Baek et al. 2021; Basavarajappa et al. 2021), where classification training sets play an important role in defining the clusters of parameter values associated with a condition.

In addition to these approaches, we argue that consideration of the underlying fundamental biophysics can refine the diagnostic value of parameter measurements. Specifically, it has recently been shown (Parker et al. 2018; Parker and Ormachea 2021), under a reasonable set of assumptions and using a model of composite material for steatotic livers, that only two measurements are sufficient to determine two key unknown quantities: the background modulus of the liver (which can vary in different disease states) and the volume percent of fat (distributed in the form of small vacuoles). The sufficient measurements are either the shear wave speed of sound and attenuation, or alternatively the ultrasound compression wave speed of sound and attenuation. While this framework has been tested using results from a human clinical trial with shear wave measurements (Parker and Ormachea 2021), there is a need for a more extensive examination of ultrasound compression wave measures for their predictive value in our model of steatosis. To accomplish that goal, this paper reviews first the key equations and assumptions leading to a quantitative model of steatosis and the solutions to two equations in two unknowns. Next, a group of reported measurements of ultrasound speed of sound and attenuation are collected with a focus on the band around 3.5 MHz, which is common in human abdominal studies. Then, these measurements or their median values are paired and assessed in both the forward model, by way of a nomogram, and the inverse model, by way of regularized optimization of the model equations. The results show reasonable agreement against magnetic resonance imaging (MRI) steatosis estimates and steatosis stages across a number of studies. These preliminary results highlight the potential for routine ultrasound quantification of liver steatosis using scanners capable of accurate speed of sound and attenuation measurements.

## THEORY

*The composite inclusion model*

The accumulation of fat in a liver is generally in the form of small spherical vesicles within the liver hepatocytes. As the vesicles grow in number, our biophysical models predict changes in scattering (Baek et al. 2020b) and biomechanical properties (Parker et al. 2018; Parker and Ormachea 2021). Recently we demonstrated (Parker and Ormachea 2021) that the complex (elastic and lossy) hepatic viscoelastic properties could be quantitatively related to the volume percent of fat. The steatotic liver is modeled as a composite material where the baseline properties are set by normal lean liver. A strong viscous or loss term is linked to fat volume fraction $V$. The general model of composite media was originally derived in a landmark paper by Christensen (1969). By imposing the principle of minimum strain energy in a deformed elastic medium, and assuming the inhomogeneities are spherical inclusions, Christensen derived bounds for the effective bulk and shear moduli for the limiting cases of the volume fraction $V$ of spheres being small, or more generally $V < 0.5$. The effective bulk modulus was derived in equation (14) of Christensen's theory (1969). Using this model, the composite liver representing simple steatosis will have a bulk modulus $B_c$ given by:

$$\frac{B_c}{B_{\text{liver}}} = 1 + \frac{3(1-v_1)\left(\frac{B_{\text{fat}}}{B_{\text{liver}}} - 1\right)V}{2(1-2v_1) + (1+v_1)\left[\frac{B_{\text{fat}}}{B_{\text{liver}}} - \left(\frac{B_{\text{fat}}}{B_{\text{liver}}} - 1\right)V\right]}. \quad (1)$$

Considering that the Poisson's ratio of the surrounding matrix approaches the incompressible limit (Fung 1981), $v_1 \approx 0.5$, and writing the frequency dependence explicitly, we find that:

$$B_c(\omega) \approx \frac{B_{\text{liver}}(\omega) \cdot B_{\text{fat}}(\omega)}{B_{\text{fat}}(\omega) + \left(B_{liver}(\omega) - B_{fat}(\omega)\right)V}, \quad (2)$$

where $B_{\text{liver}}$ represents the bulk modulus of the normal liver, $B_{\text{fat}}$ is the bulk modulus of the fat vacuoles, and $\omega$ is the radial frequency of the ultrasound waves. Equation (2) assumes a small volume fraction of fat $V$ (triglyceride-filled spherical vacuoles) contained within the viscoelastic normal liver and fat bulk modulus, $B_{\text{liver}}(\omega)$ and $B_{\text{fat}}(\omega)$, respectively.

*The composite inclusion model and its relationship with the speed of sound and attenuation parameters*

Once $B_c(\omega)$ is specified, the storage modulus and loss modulus can be obtained from the real and imaginary parts of $B_c(\omega)$, respectively. Furthermore, the complex wavenumber $\hat{k}$ is related to the bulk modulus (Blackstock 2000; Carstensen and Parker 2014) as:

$$\hat{k} = \frac{\omega}{\sqrt{\frac{B_c(\omega)}{\rho}}} = \frac{\omega}{c_l} - i\alpha, \qquad (3)$$

where $c_l$ is the longitudinal wave speed of sound, $\alpha$ is the longitudinal attenuation, and $\rho$ is the mass density (assumed to be approximately 1 g/cm$^3$ for soft tissues). In practice, the speed of sound and attenuation can be now be measured using several advanced clinical imaging platforms (Dioguardi Burgio et al. 2019; Ferraioli et al. 2021). Assuming $c_l$ and $\alpha$ have been measured accurately, it is then possible to determine $B_c(\omega)$ as:

$$B_c(\omega) = \frac{\rho \omega^2}{\left(\frac{\omega}{c_l} - i\alpha\right)^2} \qquad (4)$$

*The inverse problem approach*

The key clinical question is how $V$ can be determined experimentally. Let us define the complex bulk modulus for normal liver and fat as:

$$B_{\text{liver}}(\omega) = B_{1_{\text{Re}}} + iB_{1_{\text{Im}}}$$
$$B_{\text{fat}}(\omega) = B_{2_{\text{Re}}} + iB_{2_{\text{Im}}} \qquad (5)$$

Rewriting equation (2) and we have:

$$B_c = \frac{(B_{1_{Re}} + iB_{1_{Im}})(B_{2_{Re}} + iB_{2_{Im}})}{(B_{2_{Re}} + iB_{2_{Im}}) + ((B_{1_{Re}} + iB_{1_{Im}}) - (B_{2_{Re}} + iB_{2_{Im}}))V} \quad (6)$$

This can be separated into two equations, one real part and another imaginary part. Separating these terms, the real part of the composite $Re[B_c]$ and the imaginary part $Im[B_c]$ can be identified:

$$Re[B_c]$$
$$= \frac{B_{1_{Re}}(B_{2_{Re}}^2 + B_{2_{Im}}^2)(1-V) + (B_{1_{Re}}^2 B_{2_{Re}} + B_{1_{Im}}^2 B_{2_{Re}})V}{B_{2_{Re}}^2 + B_{2_{Im}}^2 - 2\left(B_{2_{Re}}^2 - B_{1_{Re}}B_{2_{Re}} + B_{2_{Im}}(B_{2_{Im}} - B_{1_{Im}})\right)V + \left((B_{1_{Re}} - B_{2_{Re}})^2 + (B_{1_{Im}} - B_{2_{Im}})^2\right)V^2}$$

$$Im[B_c]$$
$$= \frac{B_{1_{Im}}(B_{2_{Re}}^2 + B_{2_{Im}}^2)(1-V) + (B_{1_{Re}}^2 B_{2_{Im}} + B_{1_{Im}}^2 B_{2_{Im}})V}{B_{2_{Re}}^2 + B_{2_{Im}}^2 - 2\left(B_{2_{Re}}^2 - B_{1_{Re}}B_{2_{Re}} + B_{2_{Im}}(B_{2_{Im}} - B_{1_{Im}})\right)V + \left((B_{1_{Re}} - B_{2_{Re}})^2 + (B_{1_{Im}} - B_{2_{Im}})^2\right)V^2}$$

(7)

and let us assume that the parameters for the lossy part of the normal liver and the fat vesicles are known; that is $B_{1_{Im}}$, and both real and imaginary terms for $B_{fat}$. In addition, $B_c$ is assumed to be accurately estimated from experimental measurements at a specific frequency $\omega$ as in equation (4). In this particular case we then have two equations in two unknowns, $B_{1_{Re}}$ (real part of $B_{liver}$) and $V$ (fat volume percentage) that can be solved using numerical methods. In practice, regularization methods are employed to minimize problems of random errors in measurements or parameters that could invalidate the system of equations.

Taking the real and imaginary parts of equation (4) numerically provides two values for the left-hand side of equation (7), which can then be solved numerically for $B_{1_{Re}}$ and $V$. Numerical solution routines search through a parameter space to find the solution, in the form of a global minimum of a corresponding minimization formulation. Thus, the steps for quantifying liver fat volume fraction from ultrasound speed of sound and attenuation measurements are:

- Measure $c_l$ and $\alpha$. Calculate $B_c(\omega)$ using equation (4) at a fixed $\omega$.
- Find the real and imaginary parts of the right-hand side of equation (4).
- Substitute those into the equation (7) for $\text{Re}[B_c]$ and $\text{Im}[B_c]$ with *a prioi* known $B_{1_{\text{Im}}}$ and $B_{\text{fat}}$.
- Solve numerically for $B_{1_{\text{Re}}}$ and $V$. Also, calculate $B_{\text{liver}}$ using equation (5).

## METHODS

*Speed of sound in in vivo liver: literature review*

Conventionally, medical ultrasound systems assume a speed of sound (SoS) for transmit and receive beamforming operations. The assumed SoS is typically held constant, usually 1540 m/s for the entire image. However, due to this assumption, the ultrasound image quality may have a degradation because the different organs may have different SoS (Jaeger et al. 2015). Multiple studies have measured hepatic SoS *in vivo* as a noninvasive biomarker for liver steatosis. In this work, we reviewed different articles that reported SoS measurements for liver tissue based on four techniques: focusing, spatial coherence, compounding, and single-path transmission. The first three selected methods were recently considered the most promising categories for SoS measurements (Wang et al. 2021). The number of reviewed articles were three for each of these methods. **Table 1** gives more details about the reviewed references for SoS measurements.

**Table 1** Literature review for hepatic speed of sound and steatosis.

| | Speed of sound (m/s) | | | | |
|---|---|---|---|---|---|
| **Authors** | **S0** | **S1** | **S2** | **S3** | **Method** |
| Boozari et al. (2010) | 1575 ± 21 | *** | *** | *** | focusing |
| Napolitano et al. (2006) | 1480 | *** | *** | *** | focusing |
| Hayashi et al. (1988) | 1538 ± 29 | *** | *** | *** | focusing |
| Imbault et al. (2017) | 1557 | 1553 | 1551 | *** | spatial coherence |
| Imbault et al. (2018) | 1570 | 1510 | *** | 1470 | spatial coherence |
| Dioguardi Burgio et al. (2019) | 1570 ± 26 | 1533 ± 26 | 1511 | 1481 ± 13 | spatial coherence |
| Stähli et al. (2020) | 1564 ± 4 | *** | *** | *** | compounding |
| Robinson et al. (1982) | 1574 ± 15 | *** | *** | *** | compounding |
| Chen et al. (1987) | 1578 ± 5.4 | *** | *** | *** | compounding |
| Shigemoto et al. (2001) | 1585 | *** | *** | *** | single-path transmission measurement |
| Lin et al. (1987) | 1574 ± 10.4 | 1565 ± 8.3 | 1548 | 1538 | single-path transmission measurement |
| Bamber and Hill (1981) | 1573 | *** | *** | *** | single-path transmission measurement |

***No reported value.

*Attenuation in in vivo liver: literature review*

The attenuation coefficient (AC) measures the acoustic energy loss when an ultrasound signal passes through a medium. There are different approaches proposed by many researchers over several decades to measure the AC (Ferraioli et al. 2021). These techniques analyze the radio frequency (RF) echo signals detected by the transducer. Some of the proposed methods are the spectral shift, spectral difference, spectral log difference, and hybrid methods (Bigelow and Labyed 2013). Thus, the AC has been used as a surrogate parameter for fat liver tissue quantification. In this work, we report AC results for *in vivo* liver patients using different ultrasound clinical systems. The number of reviewed articles were: four using the 2-D attenuation imaging (ATI) system (Aplio i800, Canon Medical Systems, Otawara, Japan), three applying the attenuation parameter (ATT) system (Aloka-Arietta, Fujifilm, previously Hitachi Ltd., Japan), two using the ultrasound-guided attenuation parameter (UGAP) system (LOGIQ E9, General Electric, Schenectady, NY, USA), one using the diagnostic system (EPIQ-7G, Philips, Bothell, WA, USA), and one using the tissue attenuation imaging (TAI) system (RS85, Samsung Medison, Seoul, Korea). We were not able to extract the AC parameter from the ultrasound-derived fat fraction (UDFF) system (Acuson S3000, Siemens Healthineers, Erlangen, Germany) since this product directly reports its estimated fat fraction percentage. **Table 2** gives more details about the reviewed references for AC measurements. For the study using the Samsung system, Jeon et al. (2021) reported TAI values based on visual steatosis grades and the controlled attenuation parameter. Thus, we included the mean and standard deviation TAI values, in **Table 2**, based on both grades.

Table 2 Literature review for hepatic attenuation coefficient and steatosis.

| | Attenuation coefficient (dB/cm/MHz) | | | | |
|---|---|---|---|---|---|
| Authors | S0 | S1 | S2 | S3 | System |
| Jeon et al. (2019) | 0.58 | 0.65 | 0.73 | | ATI - Canon |
| Yoo et al. (2020) | 0.55 ± 0.08 | 0.66 ± 0.08 | 0.76 ± 0.09 | 0.76 ± 0.09 | |
| Dioguardi Burgio et al. (2020) | 0.63 ± 0.09 | 0.71 ± 0.11 | 0.87 ± 0.09 | 0.87 ± 0.09 | |
| Ferraioli et al. (2019) | 0.56 | 0.68 | 0.85 | 0.85 | |
| Cerit et al. (2020) | 0.56 | 0.62 | 0.73 | | ATT - Fujifilm/Hitachi |
| Koizumi et al. (2019) | 0.57 | 0.63 | 0.72 | 0.72 | |
| Tamaki et al. (2018) | 0.55 | 0.63 | 0.69 | 0.69 | |
| Fujiwara et al. (2018) | 0.49 | 0.56 | 0.66 | 0.66 | UGAP - GE |
| Tada et al. (2020) | 0.53 | 0.65 | 0.78 | 0.78 | |
| D'Hondt et al. (2021) | 0.48 ± 0.08 | 0.54 ± 0.03 | 0.57 ± 0.04 | 0.57 ± 0.04 | Philips |
| Jeon et al. (2021) | 0.66 ± 0.05 | 0.75 ± 0.07 | 1.11 ± 0.13 | 0.88 ± 0.09 | TAI - Samsung |
| | 0.69 ± 0.07 | 0.81 ± 0.07 | 0.85 ± 0.12 | 1.03 ± 0.13 | |

*Solution by nomogram*

A nomogram can be employed as a simplified graphical solution approach. To generate a nomogram, the forward solution is calculated from equations (2)-(4), and the resulting theoretical values of $c_l$ and $\alpha$ are plotted on a graph with contours representing regular increments of $\{V, B_{1_{Re}}\}$ values. Then, any measured pair of $\{c_l, \alpha\}$ can specify a unique point location within the contours, which provides an immediate graphical estimate of the corresponding $\{V, B_{1_{Re}}\}$. As an example, see **Figure 1**. Any patient data that falls outside of the contour ranges would indicate that there were possible errors in the measurements or that some model parameters need to be adjusted. The reader should also notice that units for AC are Np/cm at 3.5 MHz. The AC units to estimate the bulk modulus must be in Np/m at a selected frequency. The conversion from AC units reported in the literature to Np/m is possible using 1 Np ≈ 8.68 dB.

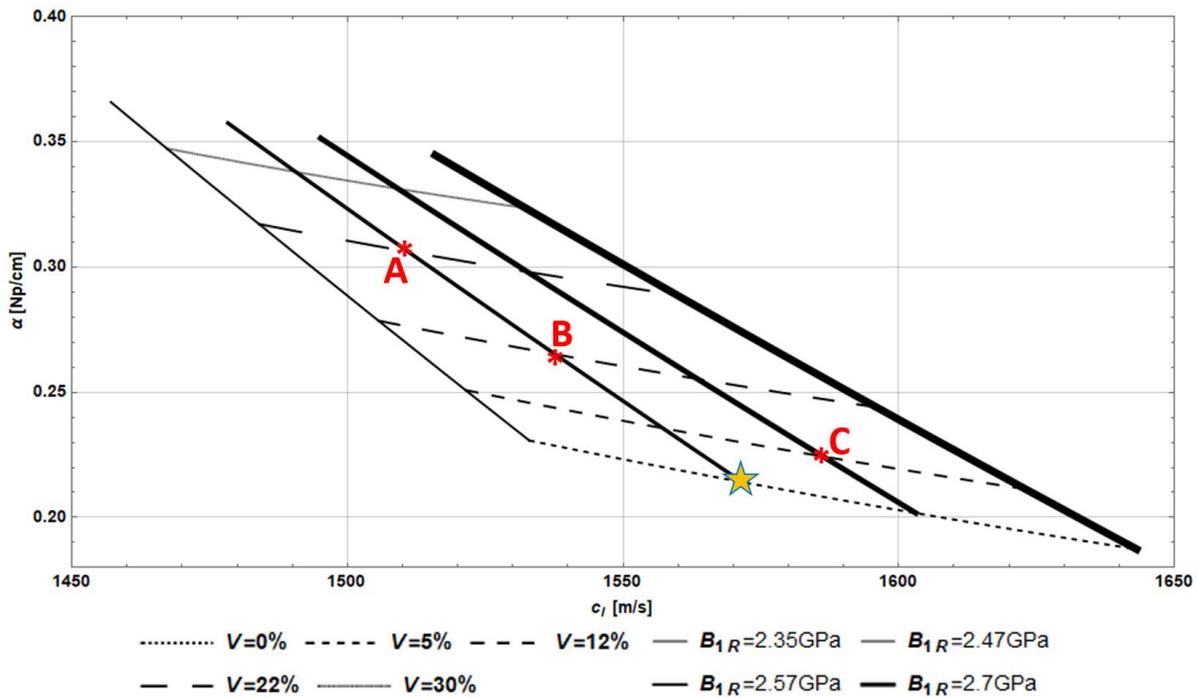

**Figure 1** Nomogram providing graphical estimates of fat volume fraction $V$ and liver matrix shear modulus $B_{1_{Re}}$, given measurements of AC (vertical axis) and SoS (horizontal axis) and assuming a frequency of 3.5 MHz. The two-dimensional parametric space is illustrated with particular values of $V$ from 0% to 30% (dashed lines), and also for particular increasing values of $B_{1_{Re}}$, (solid lines). Any measured liver values of attenuation and speed can be plotted on the nomogram to provide a graphical estimate of the $V$ and $B_{1_{Re}}$, values for that liver. For example, given three pair of $\{c_l, \alpha\}$ points A, B, and C, their corresponding $\{V, B_{1_{Re}}\}$ would be {22%, 2.47GPa}, {12%, 2.47GPa}, and {5%, 2.57GPa}, respectively. The yellow star point illustrates the assumed $\{c_l, \alpha\}$ for a normal liver with very low fat and measured at 3.5 MHz.

*Solution by the inverse problem*

Our numerical solution for $V$ and $B_{1_{Re}}$ using a minimization approach subtracts the magnitude of the terms of the next equation. This should approach zero as the correct values of $V$ and $B_{1_{Re}}$ are determined within the system of equations. The real and imaginary parts are equally weighted, although this may be varied in future research. We also limit the search parameter within practical ranges, and simulated annealing is employed to avoid the issue of entrapment within local minima. In summary, our specific routine to find the minimum solution $T$ is as follows:

<u>Min</u>

$$T = \left|Re[B'_m] - Re[B_c]\right| + \left|Im[B'_m] - Im[B_c]\right|$$

<u>s.t.</u>
(8)

$$\begin{array}{ccccc} 0.001 & < & V & < & 0.65 \\ 1.9 \text{ GPa} & < & B_{1_{Re}} & < & 2.6 \text{ GPa} \\ 0.95\, Re[B_m] & < & Re[B'_m] & < & 1.05\, Re[B_m] \\ 0.95\, Im[B_m] & < & Im[B'_m] & < & 1.05\, Im[B_m] \end{array}$$

where $B_m$ is obtained from the measured $c_l$ and $\alpha$ using equation (4) and where $B'_m$ is the approximate composite modulus entered into the equation, allowed to have a few percent variation from the measured modulus $B_m$ (due to the imprecision of measurements), and where the two unknowns are $B_{1_{Re}}$ (the real part of the liver's bulk modulus) and $V$ (the volume fraction of fat vesicles) which are linked to the composite modulus. The simulated annealing algorithm searches within constraints on the permitted values of $V$ and $B_{1_{Re}}$: $0.001 < V < 0.65$ and $1.9 \text{ GPa} < B_{1_{Re}} < 2.6 \text{ GPa}$. The upper limit, $V_h$, for $V$ was defined with one additional practical criterion, if the ratio $Im[B'_m] / Re[B'_m]$ was lower than 0.004, then $V_h = 0.4$, else $V_h = 0.65$ otherwise. In order to match the literature review data for SoS and AC, we assumed a frequency of 3.5 MHz, an imaginary part of 8.09 MPa for normal liver, and a complex bulk modulus $B_{\text{fat}}$= 1.8 GPa + $i$12.9 MPa The reasoning to select these values, for $B_{1_{\text{Im}}}$ and $B_{\text{fat}}$ respectively, is presented in the next section. After obtaining $B_{1_{\text{Re}}}$, $B_{\text{liver}}$ is calculated using equation (5) with $B_{1_{\text{Im}}} = 8.09$ MPa Using the "NMinimize" function in Mathematica (version 12.1.1.0, Wolfram, Champaign, IL, USA), the numerical solution representing the global minimum is obtained from the following command:

$$\text{NMinimize}[\{T, \text{constraints}\}, \{V, B_{1_{\text{Re}}}, \text{Re}[B'_m], \text{Im}[B'_m]\}, \text{Method} \rightarrow \text{"SimulatedAnnealing"}] \qquad (9)$$

## RESULTS

*Speed of sound*

Overall, hepatic SoS varied from 1470 m/s to 1590 m/s depending on the underlying pathology. Normal liver SoS were approximately 1570 m/s, while fatty livers showed lower SoS values. While Bamber and Hill (1981), Chen et al. (1987), and Hayashi et al. (1988) did not report SoS values for S1, S2, and S3 groups, they reported SoS values of 1547 ± 17.8 m/s, 1423 ± 34 m/s, and 1556 m/s for fatty liver tissue, respectively. **Figure 2** illustrates and summarizes the hepatic SoS values reported in the literature for normal, S0, S1, S2, and S3 steatosis stages.

*Attenuation coefficient*

Based on the literature review, the AC ranges from 0.48 to 1.11 dB/cm/MHz. The AC values for normal liver fall within the range of 0.48 to 0.69 dB/cm/MHz, whereas the AC value falls in the range of 0.54 to 0.81 dB/cm/MHz, 0.57 to 0.88 dB/cm/MHz, and 0.72 to 1.11 dB/cm/MHz for S1, S2, and S3 stages, respectively. **Figure 3** illustrates and summarizes the hepatic AC values reported in the literature for normal S0, S1, S2, and S3 steatosis stages.

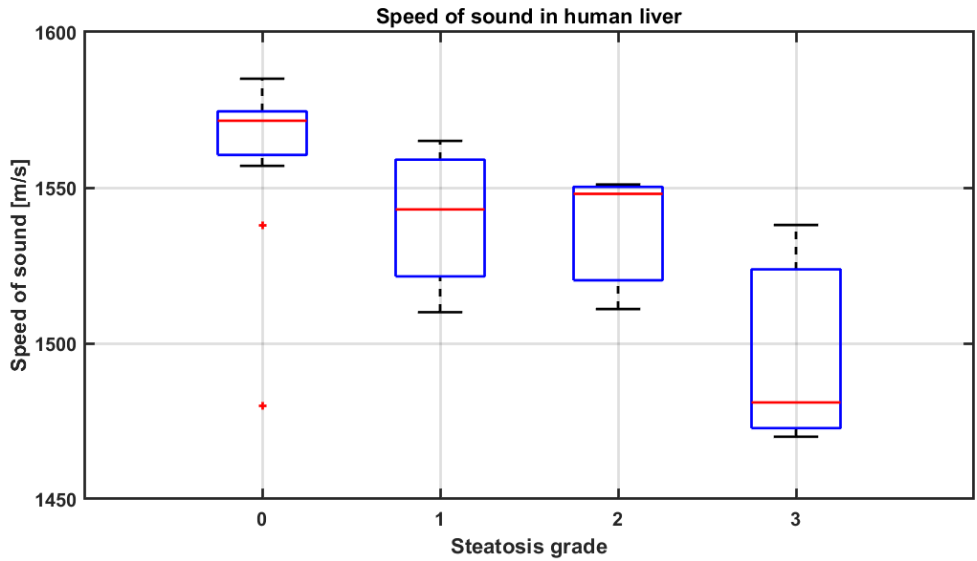

**Figure 2** Boxplot showing the median SoS values and interquartile range for each steatosis stage obtained from the literature review from different studies reporting hepatic speed of sound versus steatosis stages. Each box represents values from the lower to upper quartile (25th – 75th percentile). The middle line represents the median values. A vertical line extends from the minimum to the maximum range, separate red points represent the excluded "outside" values. The review data is presented in **Table 1**.

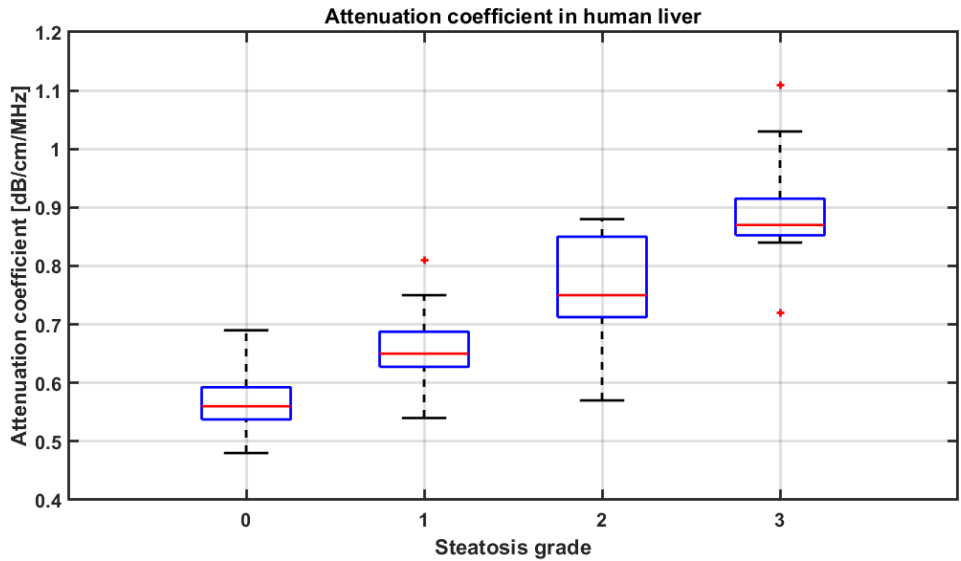

**Figure 3** Boxplot showing the median AC values and interquartile range for each steatosis stage obtained from the literature review from different studies reporting hepatic attenuation coefficient versus steatosis stages. Each box represents values from the lower to upper quartile ($25^{th}$ –$75^{th}$ percentile). The middle line represents the median values. A vertical line extends from the minimum to the maximum range, separate red points represent the excluded "outside" values. The review data is presented in **Table 2**.

*Composite modulus for normal liver and fat*

A bulk modulus for normal liver was assumed based on the median values of SoS (1570 m/s) and AC (0.55 dB/cm/MHz) obtained from the S0 (normal) groups in **Figure 1** and **Figure 2**. Then, a normal $B_{\text{liver}} = 2.46$ GPa $+ i7.56$ MPa was obtained using equation (4). Thus, $B_{1_{\text{Im}}}$ is equal to 7.56 MPa. For the fat bulk modulus case, we used a SoS of 1400 m/s based on the work by Azman and Abd Hamid (2017) determining the SoS of different types of edible oils, and we used an AC of to 1.39 dB/cm/MHz based on the studies by Chanamai and McClements (1998) and Ghosal et al. (2012) evaluating edible oils and fatty livers, respectively, and we used a density value of 0.92 g/cm$^3$ (Abe et al. 2020). Assuming these values and using equation (4), $B_{\text{fat}} = 1.8$ GPa $+ i12.9$ MPa. Therefore, we kept these values, $B_{1_{\text{Im}}}$ and $B_{\text{fat}}$, as constants for the inverse problem solution.

*Random selection of SoS and AC pairs based on literature review results*

In order to evaluate the variability within the nomogram approach and the inverse problem solution, we generated 9 random pairs of SoS and AC selected from within each interquartile range group: normal (S0), S1, S2, and S3. For example, normal liver SoS and AC values are in the range of 1540-1570 m/s and 0.4-0.7 dB/cm/MHz, respectively. Thus, we randomly selected nine SoS and AC for each range and generated nine pairs, e.g., {1565 m/s, 0.56 dB/cm/MHz} for each steatosis group. Moreover, the median SoS values reported by Dioguardio Burgio et al. (2019) and the median AC values reported by Ferraioli et al. (2019) were selected to form four additional pairs (one for each steatosis group). We selected these values because both studies compared their results against a fat fraction value that is given by the magnetic resonance imaging proton density fat fraction (MRI-PDFF) parameter. MRI-PDFF is considered an alternative, noninvasive, gold standard method for fat liver quantification and hepatic steatosis staging. Thus, we simulated 40 measurements of SoS and AC for *in vivo* liver patients that are properly bounded by clinical studies reported in the literature.

*Nomogram*

**Figure 4** shows individual pairs of SoS and AC values on the nomograph. Curves show fat volume fraction *V* at increasing values, and generally correspond to increasing grades of steatosis. *V* values correspond to similar fat cutoffs in studies that measure MRI-PDFF: $S0 \leq 6.5\%$, $S1 < 16.5\%$, $S3 < 22\%$, and $S3 \geq 22\%$ (Dioguardi Burgio et al. 2019; Ferraioli et al. 2019).

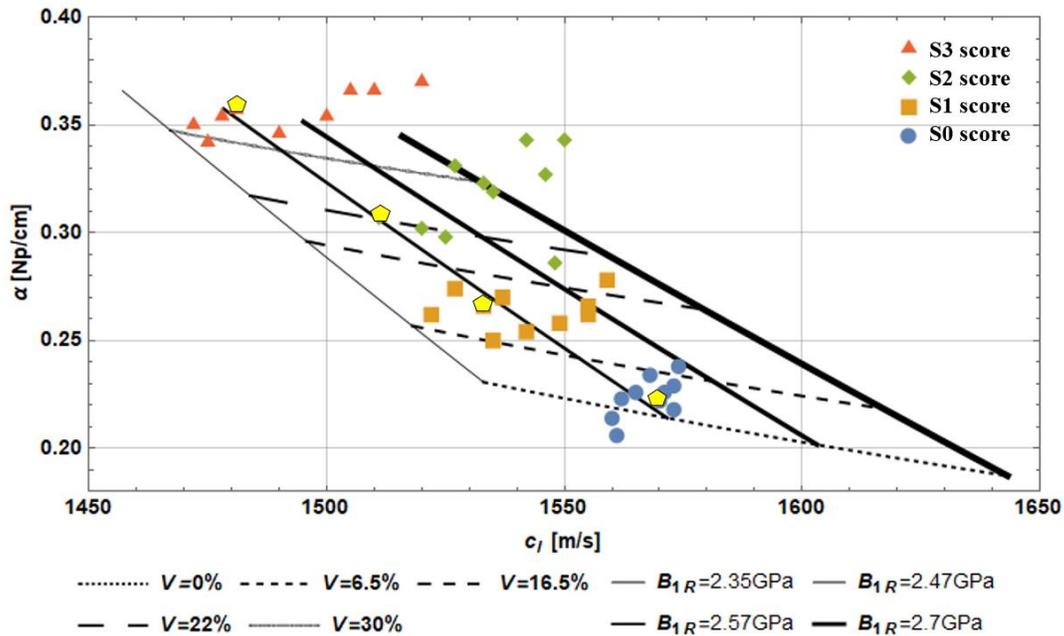

**Figure 4** Sensitivity of theory vs. experiments, showing the randomly selected $\{c_l, \alpha\}$ pair points from within steatosis scores of S0 (blue circles), S1 (orange squares), S2 (green diamonds), and S3 (red triangles). Theoretical dashed-line curves represent values of *V* equal to 0%, 6.5%, 16.5%, 22%, and 30% covering different liver bulk modulus values, $B_{1_{Re}}$, between 2.35 GPa and 2.7 GPa. The selected data are found to be stratified such that the ten cases of S3 are located above the *V* = 30% curve. Most of the S2 cases located between *V* = 16.5% and 30% curves. Most of the S1 cases are below the *V* = 16.5% curve, and most of the S0 points are below *V* = 6.5%. These *V* ranges are in agreement with the MRI-PDFF steatosis grade used in Dioguardi Burgio et al. (2019): $S0 \leq 6.5\%$, $S1 < 16.5\%$, $S2 < 22\%$, and $S3 \geq 22\%$. The yellow pentagons are the median SoS and AC values from Dioguardi Burgio et al. (2019) and Ferraoili et al. (2019), respectively. For these four $\{c_l, \alpha\}$ pair points, *V* is close to 3%, 12%, 22%, and higher than 30% with normal bulk liver $B_{1_{Re}} = 2.47$ GPa.

*Inverse problem*

**Figure 5** shows the numerical estimates from the 40 selected pairs of SoS and AC within the literature review showing the estimated volume percentage *V* of fat as a function steatosis stages S0 to S3. The steady increase in estimated *V* is observed. The median values for each group are 6% (S0), 13.5% (S1), 19.3% (S2), and 21.3% (S3). The diamond points for each boxplot are the calculated *V* using the median SoS and AC values from Dioguardio Burgio et al. (2019) and Ferraioli et al. (2019), respectively. For these four $\{c_l, \alpha\}$ pair points, *V* is equal to 7% (S0), 13% (S1), 19% (S2), and 23% (S3); these *V* values agree and correlate with the MRI-PDFF steatosis similar grades used in these two studies: S0 ≤ 6.5%, S1 < 16.5%, S3 < 22%, and S3 ≥22%.

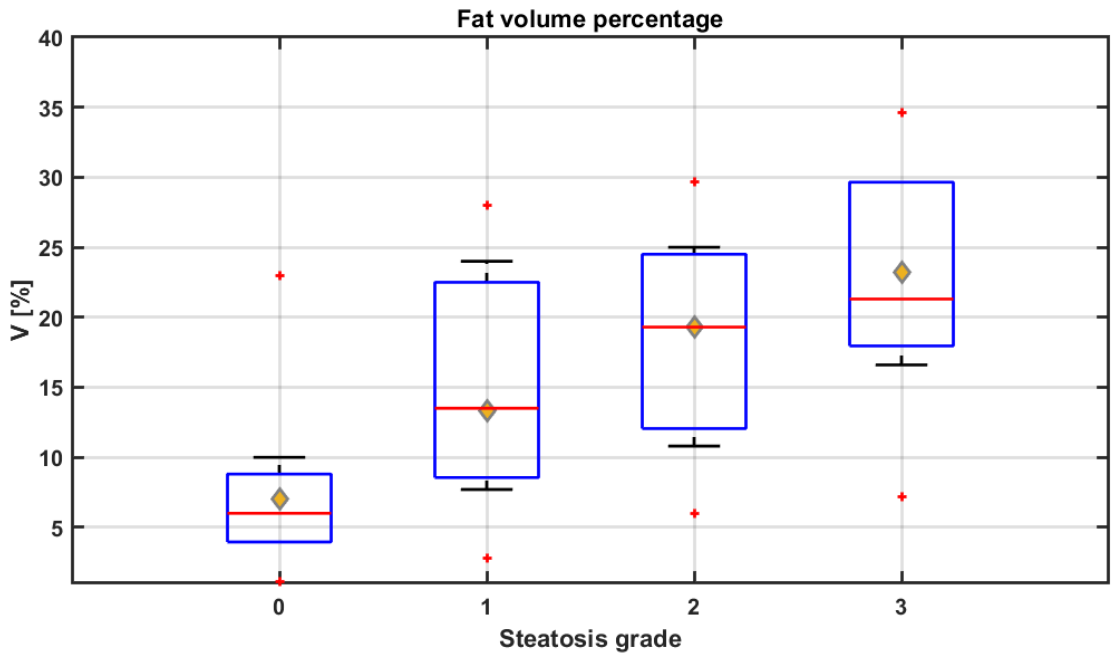

**Figure 5** Numerical solution of volume percentage fat *V* from the 40 randomly selected pairs of AC and SoS. The steady increase in estimated *V* is observed. The diamond points, at each boxplot, are the calculated *V* using the median SoS and AC values from Dioguardi Burgio et al. (2019) and Ferraioli et al. (2019), respectively. For these four $\{c_l, \alpha\}$ pair points, *V* equals to 7%, 13%, 19%, and 23%; these *V* values agree with the MRI-PDFF steatosis grade used in these two studies: S0 ≤ 6.5%, S1 < 16.5%, S3 < 22%, and S3 ≥ 22%. Separate red points represent the excluded "outside" values at each group.

# DISCUSSION

Both the nomogram approach and the inverse numerical solution approach produce reasonable agreement with independent assessments of fat. In particular, MRI-PDFF assesses the concentration of mobile triglycerides within the hepatic tissue (Caussy et al. 2018). It is an imaging biomarker that has excellent diagnostic value for assessment of hepatic fat content in patients with nonalcoholic fatty liver disease (NAFLD) (Jayakumar et al. 2019). Several clinical studies have used this measurement and compared it against histological assessments from liver biopsy, demonstrating that MRI-PDFF is highly reproducible, accurate, precise, and a reliable biomarker (Caussy et al. 2018; Ferraioli et al. 2019; Gu et al. 2019; Jayakumar et al. 2019; Stine et al. 2020; Tada et al. 2020). In addition, unlike liver biopsy, it can be used for longitudinal studies to evaluate liver fat content changes (Jayakumar et al. 2019). Moreover, MRI-PDFF has been used as an alternative reference standard measurement in clinical trials that measure AC, for example: Tada et al. (2020) and D'Hondt et al. (2021). Overall, MRI-PDFF shows a direct correlation between liver fat content and liver steatosis stage. However, clear fat percentage cutoffs and ranges for each steatotic group are not yet defined. Jayakumar et al. (2019) reported median (interquartile) MRI-PDFF ranges of 5.8% (5.2, 6.7), 14.5% (10.3, 19.2), 26.2% (18.2, 29.0), and 32.2% (26.0, 38.4), in patients with S0, S1, S2, and S3 steatosis, respectively. Jeon et al. (2019) reported steatotic MRI-PDFF cutoffs of: normal < 5%, mild 5-10 %, moderate and severe >10%. Ferraioli et al. (2021) reported MRI-PDFF cutoffs of S0 < 5%, S1 $\leq$ 16.3%, S2 < 21.6%, and S3 $\geq$ 21.6%. Similarly, Dioguardi Burgio et al. (2019) reported cutoffs as: S0 $\leq$ 6.5%, S1 $\leq$ 16.5%, S2 < 22%, and S3 $\geq$ 22 %. Further studies are needed to better define these ranges; nevertheless, it can be noted that our preliminary fat content $V$ results correlate with these fat percentage ranges. Our estimated median values are 6%, 13.5%, 19.3%, and 21.3% for groups S0, S1, S2, and S3, respectively. Furthermore, the estimated fat content, using the median values of SoS and AC from Dioguardi Burgio et al. (2019) and Ferraioli et al. (2021) are in good agreement with the MRI-PDFF threshold values used in both studies. Thus, our approach could be an alternative method that noninvasively assess the *in vivo* hepatic fat content using ultrasound systems that are less expensive, more portable, and more widely available around the world than MRI systems.

Limitations of this study and the approach include the need for higher precision in the measurements of speed of sound and attenuation, and also the determination of model parameters. Since many efforts are ongoing to improve ultrasound measurements and their implementation on clinical scanners, it is anticipated that these will become more accurate and precise over time. However, more physical and mechanical measurements may be required to independently confirm the baseline values of bulk moduli used in the model. The variability within individuals of the bulk moduli of the fat/oil component is an area requiring further research. Furthermore, the model predicts that fibrosis within a steatotic liver is an important cofactor by means of changing the bulk modulus of the liver, however this needs further study with independent confirmation of the degree of influence. Also, the effects of other cofactors such as inflammation or high blood pressure have not been incorporated.

It is interesting to compare these results using ultrasound waves and bulk moduli against those obtained recently using shear waves and shear moduli (Ormachea and Parker 2021). The composite model and nomograms for each case have some similarities, but the precision required to accurately determine fat volume percentage *V* from the measured parameters differs. This is a larger subject which will require further refinement, but it is possible to speculate that a clinical scanner capable of four measurements (speed and attenuation from both ultrasound and shear waves) would be able to improve the final estimation of fat content by utilizing both sets of solutions. This remains for future work as measurement capabilities increase in clinical scanners.

## CONCLUSION

An analysis of the composite model of hepatic steatosis was performed, and a fat quantification of liver was achieved using the longitudinal (compressional) speed of sound and attenuation coefficient with results in good agreement with reported values of MRI-PDFF. The composite model approach has the potential to be implemented in commercial clinical systems for rapid steatosis staging. Thus, a fat quantification could be achieved in a safe, low-cost ultrasound system. The obtained estimate of liver fat volume percentage $V$ could be used as a quantitative biomarker for assessing and monitoring liver fat concentration for hepatic steatosis.

## ACKNOWLEDGMENTS

This work was supported by the Hajim School of Engineering and Applied Sciences and the Department of Electrical and Computer Engineering at the University of Rochester.

## CONFLICT OF INTEREST STATEMENT

The authors and the University of Rochester have patent applications pending which may pertain to some of the subject matter discussed herein.